\documentclass[prl,twocolumn,superscriptaddress,preprintnumbers,amsmath,amssymb]{revtex4-2}
\usepackage{graphicx}
\usepackage{dcolumn}
\usepackage[dvipsnames]{xcolor}
\usepackage{amsmath,amsbsy,amssymb,scalefnt}
\usepackage{mathrsfs}
\usepackage{bm}
\usepackage{booktabs}
\usepackage[T1]{fontenc}
\usepackage{amsfonts}
\usepackage{colortbl}
\usepackage{textcomp}
\usepackage{mathtools}
\usepackage{hyperref}
\usepackage{physics}

\hypersetup{ pdfnewwindow=true, colorlinks=true, linkcolor=NavyBlue, anchorcolor=NavyBlue, citecolor=blue!20!red, filecolor=blue, menucolor=blue, urlcolor=blue}
\begin{document}

\title{Optical activation of nonlinear Hall effect\\ in topological insulators with warped Fermi surface}
\author{Mohammad Shafiei}
\affiliation{COMMIT, Department of Physics \& NANOlight Center of Excellence, University of Antwerp, Groenenborgerlaan 171, B-2020 Antwerp, Belgium}
\author{Farhad Fazileh}
\thanks{\textcolor{blue}{Deceased.}}
\affiliation{Department of Physics, Isfahan University of Technology, Isfahan 84156-83111, Iran}
\author{Milorad V. Milo\v{s}evi\'c}
\email{milorad.milosevic@uantwerpen.be}
\affiliation{COMMIT, Department of Physics \& NANOlight Center of Excellence, University of Antwerp, Groenenborgerlaan 171, B-2020 Antwerp, Belgium}

\date{\today}

\begin{abstract}
Topological insulators (TIs) with hexagonally warped Fermi surface are natural platforms for the nonlinear Hall effect, as warping breaks inversion symmetry while preserving time-reversal symmetry (TRS). Here we show that this inversion breaking alone is insufficient: although warping generates a strongly anisotropic Berry curvature, the preserved threefold rotational symmetry forces the equilibrium Berry curvature dipole (BCD) to vanish identically. We demonstrate that linearly polarized light removes this symmetry obstruction: in the off-resonant Floquet regime, it lowers the rotational symmetry while preserving TRS, thereby generating a finite BCD whose magnitude, orientation, and sign are continuously tunable by the light intensity and polarization. For realistic Bi$_2$Te$_3$ parameters, we show that the induced BCD reaches $\sim$0.03~nm, yielding microampere-scale nonlinear Hall currents under experimentally accessible conditions. Our results therefore establish Floquet symmetry engineering as a route to activating the symmetry-forbidden nonlinear transport on TI surfaces without breaking the TRS.
\end{abstract}

\maketitle

\paragraph{Introduction.}
The nonlinear Hall effect (NHE) enables a transverse electrical response in time-reversal symmetric materials without an external magnetic field~\cite{sodemann2015quantum,nagaosa2010anomalous,xiao2010berry}. Its intrinsic contribution is governed by the Berry curvature dipole (BCD), the first momentum space moment of the Berry curvature over the Fermi surface~\cite{sodemann2015quantum,liu2025quantum}. This geometric mechanism has motivated extensive efforts to realize nonlinear rectification, frequency conversion, and electrically controllable transverse transport in quantum materials~\cite{ma2019observation,du2021nonlinear}.

Inversion symmetry (IS) breaking is necessary for a finite BCD, but it is not sufficient. Since the BCD transforms as an in-plane vector, proper rotations of order three or higher force the BCD to vanish. Intrinsic NHE has therefore been observed mainly in low-symmetry Weyl and Dirac materials, including WTe$_2$ and related compounds, whose point groups permit a nonzero dipole~\cite{kang2019nonlinear,ma2019observation}. Extending this physics to structurally simpler and experimentally mature platforms remains challenging and commonly requires strain, circularly polarized light, gating, or proximity to a band topology transition~\cite{du2018band,zhang2018berry,du2021nonlinear,qin2024light}.

Topological insulators (TIs), having hexagonally warped Fermi surface, offer an appealing alternative. The cubic warping term of their Hamiltonian breaks IS, deforms the surface Fermi contour into the characteristic snowflake shape, and generates a pronounced Berry curvature texture~\cite{fu2009hexagonal}. These features suggest that warped TI surface states should naturally exhibit an intrinsic NHE. The central result of this work is that this inference is incomplete: warping prepares the required Berry curvature landscape, but the remaining threefold rotational symmetry prevents that landscape from acquiring a finite dipole moment, hence NHE is not possible.

In this work, we establish a symmetry based route to activating the NHE on hexagonally warped TI surface states. We first derive the effective Floquet Hamiltonian for a surface state driven by off-resonant linearly polarized light (LPL) and obtain the corresponding Berry curvature of the dressed bands. We then show that, despite the pronounced Berry curvature texture generated by hexagonal warping, the equilibrium BCD vanishes identically as a consequence of the preserved threefold rotational symmetry. Next, we demonstrate that LPL lifts this symmetry constraint through a time-reversal symmetric momentum-dependent Floquet mass, thereby activating a finite BCD. We systematically investigate its dependence on the warping strength, Fermi energy, field amplitude, and polarization direction, revealing continuous control of its magnitude, orientation, and sign. Finally, using parameters representative of Bi$_2$Te$_3$, we estimate the resulting nonlinear Hall signal and discuss experimentally accessible signatures of the proposed optical mechanism for symmetry engineering.

\paragraph{Model and Floquet Hamiltonian.}
We consider the (001) surface of a three dimensional TI described at low energy by the warped Dirac Hamiltonian~\cite{fu2009hexagonal,liu2010model,akzyanov2019bulk}
\begin{equation}
\mathcal{H}_0(\mathbf{k})=
 v_{\mathrm F}(k_y\sigma_x-k_x\sigma_y)
 +\lambda_h(k_x^3-3k_xk_y^2)\sigma_z,
\label{eq:H0}
\end{equation}
where $v_{\mathrm F}$ is the Fermi velocity, $\lambda_h$ is the hexagonal warping strength, and $\bm\sigma$ denotes the Pauli matrices. Unless otherwise stated, we use the Bi$_2$Te$_3$ values $v_{\mathrm F}=0.255\,\mathrm{eV\,nm}$ and $\lambda_h=0.25\,\mathrm{eV\,nm^3}$ inferred from ARPES and $k\cdot p$ modeling~\cite{chen2009experimental,fu2009hexagonal,liu2010model}.

The effect of a normally incident linearly polarized field is included through the time-dependent momentum shift
\begin{equation}
\mathbf{k}\rightarrow \mathbf{k}-\mathbf{a}(t),\qquad
\mathbf{a}(t)=A_0\cos(\omega t)(\cos\theta,\sin\theta),
\label{eq:drive}
\end{equation}
where $\theta$ is the polarization angle and $A_0=eE_0/(\hbar\omega)$ has units of inverse length. This notation absorbs the factor $e/\hbar$ associated with minimal coupling into the field amplitude and avoids introducing a second vector potential convention.

In the off-resonant high-frequency regime, the leading term of the Floquet--Magnus expansion is the cycle average~\cite{mananga2016floquet,sen2021analytic}
\begin{equation}
\mathcal{H}_{\mathrm{eff}}^{(0)}=
\frac{1}{T}\int_0^T dt\,\mathcal{H}_0[\mathbf{k}-\mathbf{a}(t)],
\label{eq:floquet_average}
\end{equation}
with higher-order corrections controlled by inverse powers of $\omega$~\cite{kitagawa2011transport,shafiei2025light,qin2024light}. Averaging the cubic warping term using $\langle\cos\omega t\rangle=\langle\cos^3\omega t\rangle=0$ and $\langle\cos^2\omega t\rangle=1/2$ yields
\begin{equation}
\mathcal{H}_{\mathrm{eff}}=
\mathcal{H}_0+
\frac{3\lambda_hA_0^2}{2}
(\cos2\theta\,k_x-\sin2\theta\,k_y)\sigma_z.
\label{eq:Heff}
\end{equation}
Equation~\eqref{eq:Heff} is the zeroth order Floquet Hamiltonian retained throughout this work. The Dirac contribution produces no static correction under the cycle average, whereas the cubic warping term generates an additional mass contribution linear in momentum. Its axis is locked to $2\theta$, making the light polarization a direct control parameter for the point group symmetry of the dressed bands.

\begin{figure}[t]
    \centering
    \includegraphics[width=0.97\linewidth]{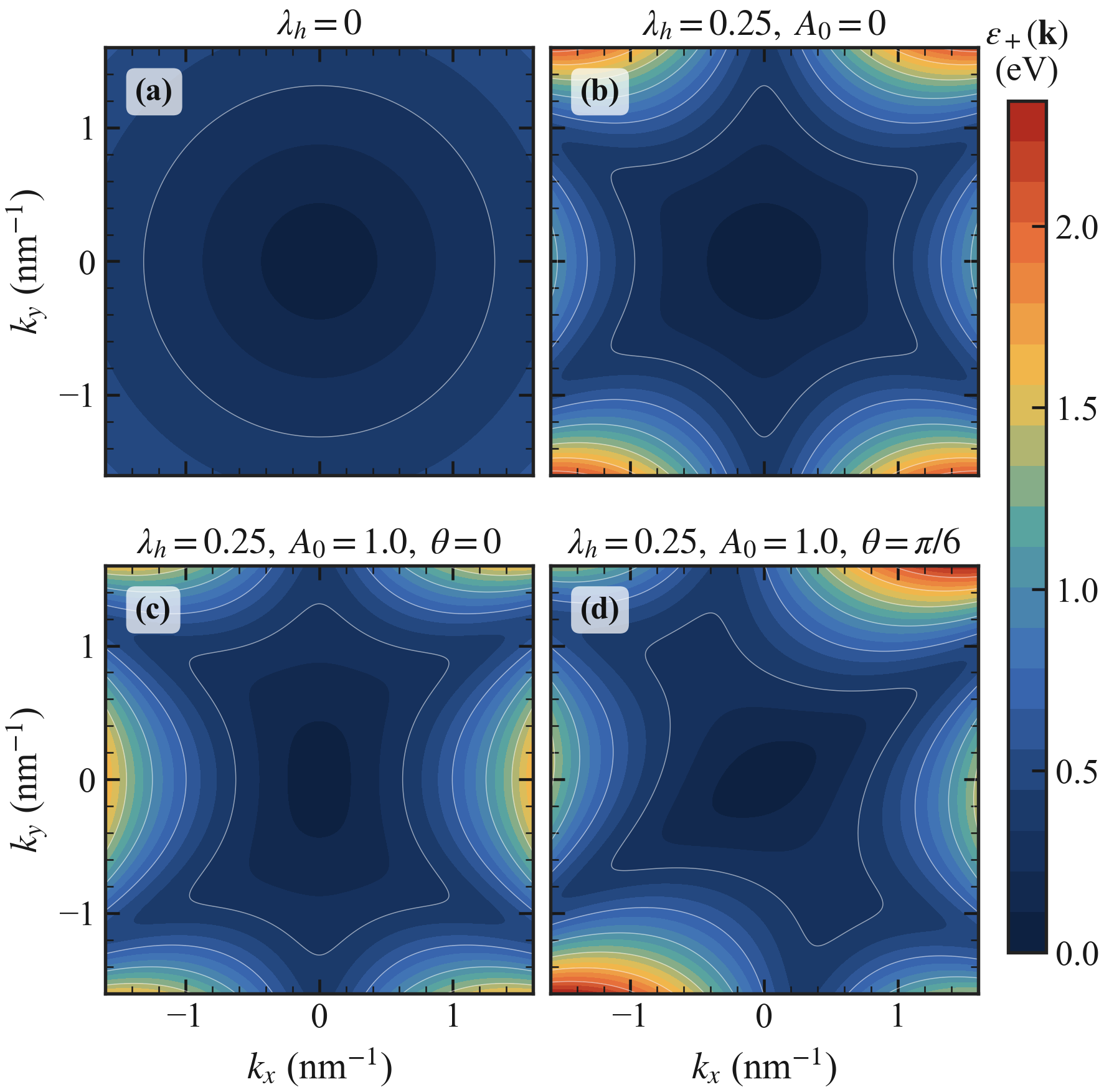}
    \caption{Evolution of the conduction band dispersion with warping and illumination. (a) Isotropic Dirac cone for $\lambda_h=0$. (b) Hexagonally warped Fermi surface in absence of illumination, retaining $C_{3v}$ symmetry. (c),(d) Photon dressed contours for $A_0=1.0\,\mathrm{nm}^{-1}$ and polarization angles $\theta=0$ and $\pi/6$, respectively. The optical field removes the threefold equivalence of the warped lobes and selects a polarization dependent anisotropy axis.}
    \label{fig:fig1}
\end{figure}

It is convenient to write Eq.~\eqref{eq:Heff} as $\mathcal{H}_{\mathrm{eff}}=\mathbf{d}(\mathbf{k})\cdot\bm\sigma$, where
\begin{equation}
\mathbf{d}(\mathbf{k})=
\big(v_{\mathrm F}k_y,-v_{\mathrm F}k_x,M(\mathbf{k})\big),
\end{equation}
and
\begin{equation}
M(\mathbf{k})=
\lambda_h(k_x^3-3k_xk_y^2)
+\frac{3\lambda_hA_0^2}{2}
(\cos2\theta\,k_x-\sin2\theta\,k_y).
\label{eq:mass}
\end{equation}
The quasistatic band energies are therefore
\begin{equation}
\varepsilon_{\pm}(\mathbf{k})=
\pm\sqrt{v_{\mathrm F}^2(k_x^2+k_y^2)+M(\mathbf{k})^2}.
\label{eq:bandstructure}
\end{equation}
Figure~\ref{fig:fig1} illustrates the corresponding conduction band contours. Without warping the dispersion is isotropic. At finite $\lambda_h$ and $A_0=0$, the contour develops the familiar threefold symmetric snowflake form. Optical dressing distorts this pattern along a polarization controlled axis, visibly removing the threefold equivalence of the lobes.

\paragraph{Berry curvature and nonlinear Hall geometry.}
For a two-band Hamiltonian $\mathbf{d}\cdot\bm\sigma$, the Berry curvature is
\begin{equation}
\Omega_{\pm}(\mathbf{k})=
\mp\frac{1}{2}
\frac{\mathbf{d}\cdot
(\partial_{k_x}\mathbf{d}\times\partial_{k_y}\mathbf{d})}
{|\mathbf{d}|^3},
\label{eq:general_BC}
\end{equation}
where the overall band sign follows the chosen eigenstate convention. Up to this conventional band sign, direct evaluation of Eq.~\eqref{eq:Heff} gives
\begin{equation}
\Omega_{\pm}(\mathbf{k})=
\pm\frac{v_{\mathrm F}^2\lambda_h
(k_x^3-3k_xk_y^2)}
{\left[v_{\mathrm F}^2(k_x^2+k_y^2)+M(\mathbf{k})^2\right]^{3/2}}.
\label{eq:BC}
\end{equation}
A notable feature of Eq.~\eqref{eq:BC} is that the light-induced term cancels from the numerator. This follows from the homogeneity of the linear contribution to $M(\mathbf{k})$: in the combination $M-\mathbf{k}\cdot\nabla_{\mathbf{k}}M$ entering the two-band curvature, every term linear in momentum vanishes, whereas the cubic warping term survives. The optical field therefore does not create Berry curvature from an otherwise curvature-free Dirac cone. Instead, it reshapes the dispersion and the denominator of Eq.~\eqref{eq:BC}, redistributing the existing warping induced curvature over momentum space.

The quantity relevant to the intrinsic NHE is not the local Berry curvature itself but its first moment over the occupied states. In two dimensions, the BCD can be written as
\begin{equation}
\rm{BCD}_\alpha=
-\sum_n\int\frac{d^2k}{(2\pi)^2}
\left(\partial_{k_\alpha}\varepsilon_n\right)
\Omega_n(\mathbf{k})
\frac{\partial f_0}{\partial\varepsilon_n},
\label{eq:BCD}
\end{equation}
where $\alpha=x,y$; $n$ labels the bands, and $f_0$ is the equilibrium Fermi distribution. Equation~\eqref{eq:BCD} makes the Fermi surface character of the response explicit: $-\partial f_0/\partial\varepsilon$ restricts the integral to the thermal window around the chemical potential. A strongly anisotropic Berry curvature texture can therefore coexist with a vanishing BCD whenever symmetry forces the velocity weighted first moment to cancel.

\begin{figure}[t]
    \centering
    \includegraphics[width=0.97\linewidth]{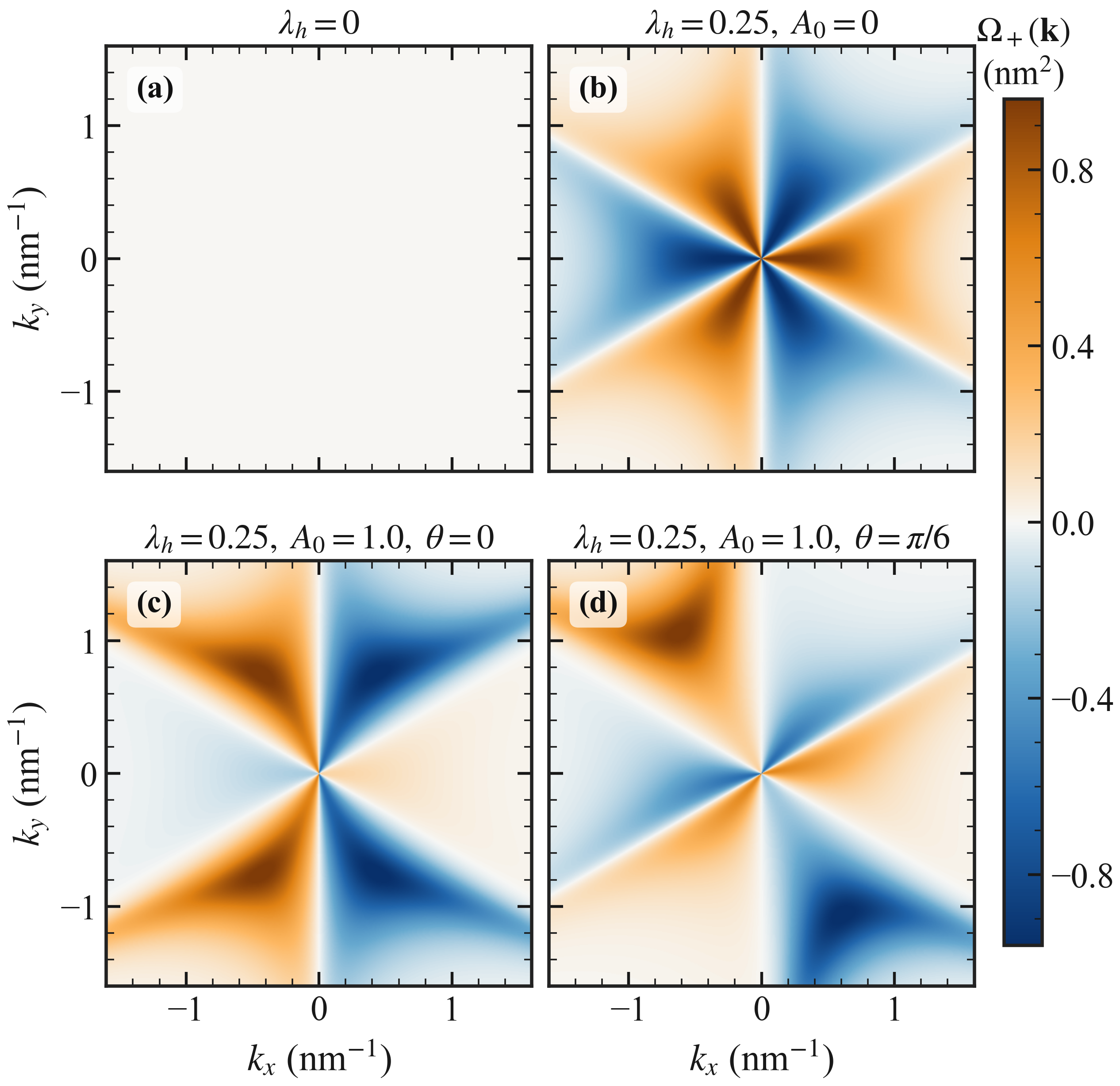}
    \caption{Conduction band Berry curvature. (a) The unwarped cone carries no warping-induced curvature. (b) Hexagonal warping produces a pronounced but $C_{3v}$-symmetric six-lobe texture. (c),(d) LPL redistributes the curvature through the dressed band denominator, making symmetry-related lobes contribute unequally to the first moment in Eq.~\eqref{eq:BCD}.}
    \label{fig:fig2}
\end{figure}

Figure~\ref{fig:fig2} displays the conduction band curvature for the same parameters as Fig.~\ref{fig:fig1}. In equilibrium, the six-lobe texture is locally large but arranged in symmetry related sectors. Under LPL, the lobe positions and weights become inequivalent because the optical mass changes the energy denominators sampled at a fixed Fermi level. This redistribution is the microscopic origin of the light-activated BCD, as derived below.

\paragraph{Symmetry protected absence of the equilibrium NHE.}
With the geometric quantities now defined, the equilibrium cancellation follows directly from point group symmetry. For $A_0=0$, the Hamiltonian is invariant under a $2\pi/3$ rotation. Indeed,
\begin{equation}
k_x^3-3k_xk_y^2=\mathrm{Re}\left[(k_x+ik_y)^3\right]
\end{equation}
is unchanged when $k_x+ik_y\rightarrow e^{i2\pi/3}(k_x+ik_y)$. Consequently, both $\varepsilon_n(\mathbf{k})$ and $\Omega_n(\mathbf{k})$ transform covariantly under the threefold rotation $R_3$.

Applying $R_3$ to Eq.~\eqref{eq:BCD} and changing the integration variable from $\mathbf{k}$ to $R_3\mathbf{k}$ gives
\begin{equation}
\mathbf{BCD}=R_3\,\mathbf{BCD}.
\label{eq:D_rotation}
\end{equation}
No nonzero in-plane vector is invariant under a $120^\circ$ rotation; hence
\begin{equation}
\mathbf{BCD}\,(A_0=0)=\mathbf{0}
\qquad
\text{for all }\lambda_h,\ E_F,\ \text{and }T.
\label{eq:zero_theorem}
\end{equation}
This result is stronger than the statement that the integrated Berry curvature vanishes. The curvature can be highly anisotropic and locally large, as in Fig.~\ref{fig:fig2}(b), while its velocity weighted dipole remains exactly zero. Thus, hexagonal warping supplies the Berry curvature texture required for a response but does not remove the rotational symmetry that forbids its first moment.

Table~\ref{tab:symmetry_analysis} summarizes this hierarchy. The equilibrium warping term breaks IS but leaves $C_{3v}$ intact, so the BCD remains forbidden. The optical correction preserves TRS but lowers the rotational symmetry and thereby removes the prohibition. This distinction is the main principle of the nonlinear response studied here.

The reduction of symmetry depends on the polarization. For generic $\theta$, the linear term in Eq.~\eqref{eq:mass} removes all in-plane point group operations and yields $C_1$. At mirror aligned polarizations, one reflection can survive, giving $C_s$. For example, at $\theta=0$, $M(k_x,k_y)$ is even under $k_y\rightarrow-k_y$ transformation. Equation~\eqref{eq:BCD} then permits $\rm{BCD}_x$ but forces $\rm{BCD}_y=0$, because the $y$-directed velocity factor is odd under the surviving mirror. Away from mirror symmetric directions, both components are generally allowed.

\begin{table}[b]
\caption{Symmetry content of the effective Hamiltonian and its consequence for the BCD. The decisive distinction is between inversion breaking, already present in equilibrium, and the rotational symmetry breaking induced by LPL.}
\label{tab:symmetry_analysis}
\centering
\begin{tabular}{@{}p{2.55cm}p{2.15cm}p{3.25cm}@{}}
\toprule
Contribution & Preserved symmetry & Consequence \\
\midrule
Dirac term & TRS, IS, $C_{3v}$ & No intrinsic BCD \\
\addlinespace[2pt]
Hexagonal warping & TRS, $C_{3v}$ & IS is broken, but the BCD remains forbidden by $C_3$ \\
\addlinespace[2pt]
Floquet correction & TRS; $C_s$ or $C_1$ & Rotational protection is lifted and a finite BCD is allowed \\
\bottomrule
\end{tabular}
\end{table}

\begin{figure}[t]
    \centering
    \includegraphics[width=0.86\linewidth]{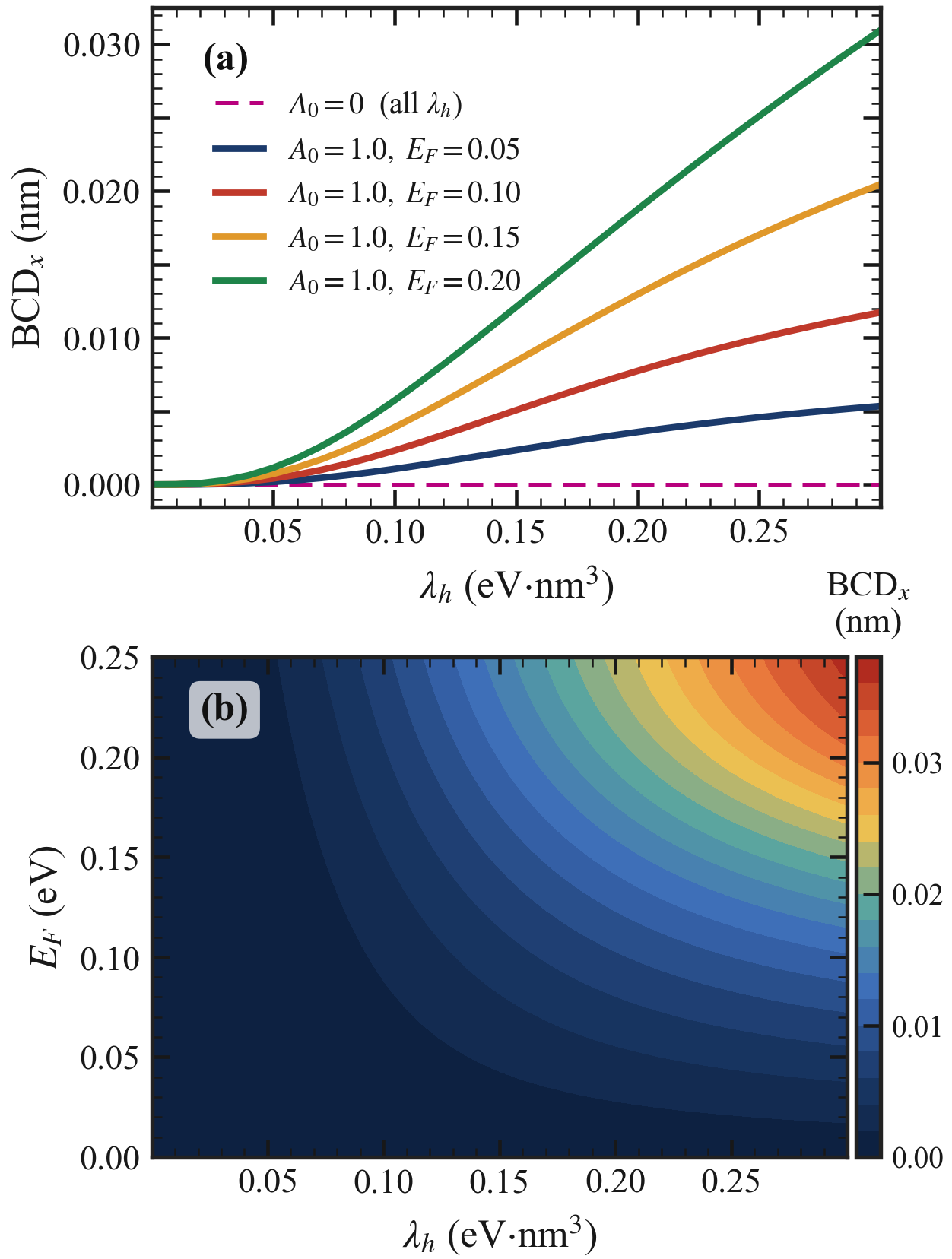}
    \caption{Activation of the Berry curvature dipole. (a) $BCD_x$ versus warping strength. The equilibrium response vanishes for all $\lambda_h$, whereas LPL produces a finite dipole that disappears in the unwarped limit. (b) $BCD_x$ in the $(\lambda_h,E_F)$ plane for $A_0=1.0\,\mathrm{nm}^{-1}$ and $\theta=0$, showing the combined dependence on curvature strength and Fermi surface geometry.}
    \label{fig:fig3}
\end{figure}

\paragraph{Optical activation of the Berry curvature dipole.}
We evaluate Eq.~\eqref{eq:BCD} numerically at $T=100\,\mathrm{K}$ using adaptive integration in polar coordinates. Figure~\ref{fig:fig3}(a) verifies Eq.~\eqref{eq:zero_theorem}: the equilibrium dipole remains zero over the entire range of $\lambda_h$. At finite illumination, $\rm{BCD}_x$ becomes nonzero and increases with the warping strength. This behavior reflects the complementary roles of the two ingredients. The optical field removes the rotational cancellation, while $\lambda_h$ supplies the Berry curvature itself; consequently, the response vanishes again as $\lambda_h\rightarrow0$. Figure~\ref{fig:fig3}(b) maps $\rm{BCD}_x$ in the $(\lambda_h,E_F)$ plane. The strongest response occurs where the warped Fermi contour samples the largest imbalance among the curvature lobes, reaching approximately 0.036~nm in the explored parameter range.

\begin{figure*}[t]
    \centering
    \includegraphics[width=0.97\linewidth]{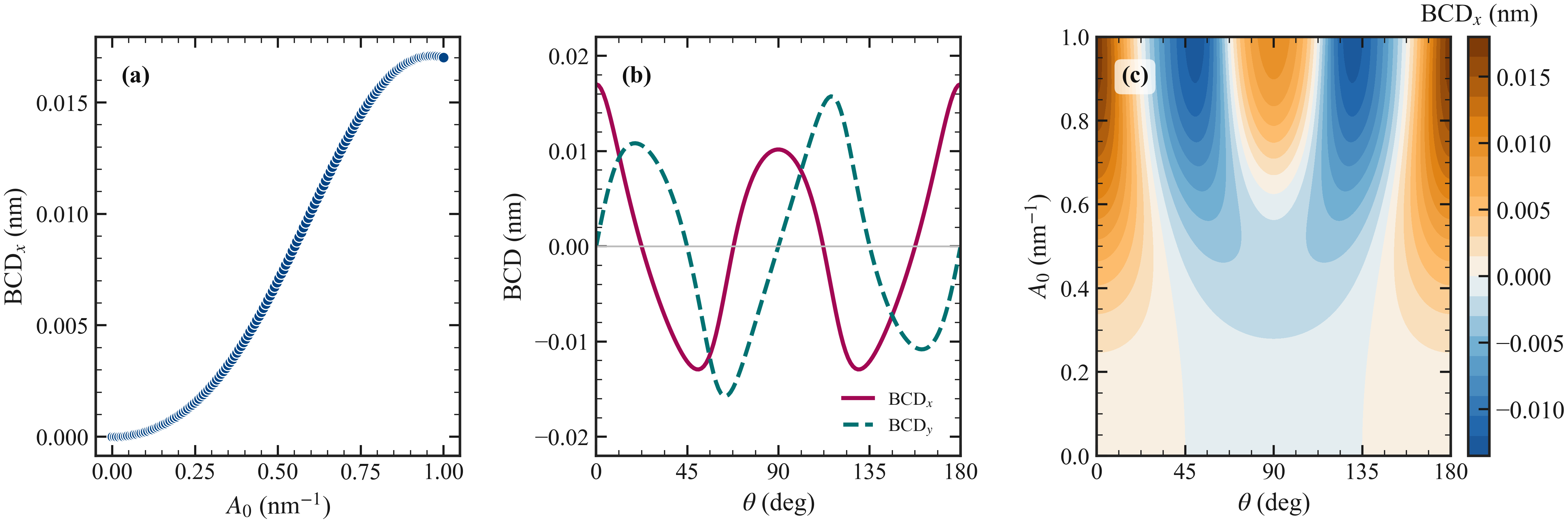}
    \caption{Optical control of the BCD for Bi$_2$Te$_3$ at E$_F$ = 0.15 eV. (a) $\rm{BCD}_x$ versus $A_0$ at $\theta=0$, displaying the expected quadratic onset. (b) $\rm{BCD}_x$ and $\rm{BCD}_y$ versus polarization angle at $A_0=1.0\,\mathrm{nm}^{-1}$. The transverse component $\rm{BCD}_y$ vanishes at mirror symmetric orientations. (c) Joint $(\theta,A_0)$ map of $\rm{BCD}_x$, demonstrating continuous amplitude control and polarization-driven sign reversal.}
    \label{fig:fig4}
\end{figure*}

The optical control parameters are explored in Fig.~\ref{fig:fig4} for E$_F$ = 0.15 eV. For small amplitude of illumination, the Floquet correction in Eq.~\eqref{eq:Heff} scales as $A_0^2$, and the induced dipole correspondingly starts with an approximately quadratic dependence. At larger amplitudes, the response becomes nonperturbative in the dressed band geometry because the same optical mass modifies both the velocity and the curvature denominator. The angular dependence is likewise richer than a single harmonic: the $2\theta$ structure of the optical term interferes with the threefold warping texture, producing alternating extrema and sign reversals. At $\theta=0^\circ,90^\circ$, and $180^\circ$, the surviving mirror symmetry forces $\rm{BCD}_y=0$; between these directions, both components are finite. The full $(\theta,A_0)$ map in Fig.~\ref{fig:fig4}(c) demonstrates reversible control over the magnitude and sign of the nonlinear Hall response without changing the chemical potential or the underlying material.

Within the relaxation time approximation, the second-order low-frequency Hall conductivity associated with $BCD_x$ is~\cite{sodemann2015quantum}
\begin{equation}
\chi_{yxx}=\frac{e^3\tau}{2\hbar^2}BCD_x,
\qquad j_y^{(2)}=\chi_{yxx}E_x^2 .
\label{eq:chi}
\end{equation}
Taking $\tau=0.5\,\mathrm{ps}$, $BCD_x=0.03\,\mathrm{nm}$, and $E_x=30$--$100\,\mathrm{kV/cm}$ yields approximately $25$--$277\,\mathrm{A/m}$ of sheet current density. For a micron scale Hall bar this corresponds to a transverse current in the microampere range. The signal can be isolated through second harmonic lock-in detection and distinguished from equilibrium backgrounds by modulating either the optical intensity or the polarization angle.

The angular response offers an additional experimental fingerprint. Rotating the polarization reverses the sign of the dipole at fixed carrier density, whereas an equilibrium nonlinear contribution from a static structural asymmetry would remain locked to the crystal axes. Simultaneous measurements of the longitudinal second harmonic response and the transverse Hall voltage could therefore separate optical symmetry breaking from heating or photoconductive effects. Since the mechanism does not require a photoinduced gap or a change of topological phase, it should also be less sensitive to fine tuning of the driving frequency than the resonant or transition-based Floquet proposals.

The present mechanism clarifies the distinct roles of warping and optical driving. Warping is indispensable because it generates the momentum space Berry curvature texture, but its native $C_{3v}$ arrangement has no dipole moment. LPL acts not as the source of curvature but as a symmetry selective weighting field that converts this texture into a nonzero first moment. This separation between curvature generation and dipole activation provides a useful design principle for other nonlinear transport platforms: a material may host a large latent Berry curvature distribution while exhibiting no NHE until a weak external perturbation lowers the relevant rotation symmetry.

The response should persist at finite temperature as long as the thermal broadening does not average over regions with opposite dipole contributions. Moderate nonmagnetic disorder enters primarily through the relaxation time and broadening of the Fermi surface, while the symmetry allowed character of the effect remains unchanged. Material optimization should therefore favor surfaces with strong hexagonal warping, long transport lifetimes, and a chemical potential intersecting the most anisotropic portion of the surface band. Bi$_2$Te$_3$ is a natural candidate, but the same principle applies to other warped Bi-chalcogenide materials and to engineered Dirac systems with analogous cubic anisotropy.

\paragraph{Conclusion.}
In conclusion, hexagonal warping does not by itself generate an intrinsic nonlinear Hall effect on a surface state of a topological insulator. Although it breaks inversion symmetry and produces a pronounced Berry-curvature texture, the residual threefold rotational symmetry forces the equilibrium Berry-curvature dipole to vanish exactly. Linearly polarized light supplies the necessary symmetry breaking through a time-reversal-preserving Floquet mass, activating a BCD whose magnitude, direction, and sign are optically programmable. The predicted response reaches experimentally measurable values for realistic Bi$_2$Te$_3$ parameters and requires neither magnetic order nor a photoinduced topological transition. Beyond its fundamental significance, this provides a practical route to dynamically controlling nonlinear transverse transport, with potential relevance for optical rectification, frequency conversion, and polarization-sensitive signal processing. More broadly, our results establish a general design principle: latent nonlinear responses in materials with symmetry-compensated Berry-curvature textures can be activated by selectively lifting the obstructing crystal symmetry, with illumination providing a particularly versatile route.

\paragraph{Acknowledgments.}
We dedicate this work to the memory of Prof. Farhad Fazileh, whose contributions were essential to this study. He passed away before the submission of the manuscript. This research was supported by the Research Foundation--Flanders (FWO-Vlaanderen), the FWO--FNRS EOS project ShapeME, and the Special Research Funds (BOF) of the University of Antwerp.

\bibliography{bibliography}

\end{document}